# The role of molecular structure on the microscopic thermodynamics: unveiling with Femtosecond Optical Tweezers

Ajitesh Singh[a,¤], Dipankar Mondal[a,b,¤], Krishna Kant Singh[a], Deepak Kumar[a] and Debabrata Goswami[a,c]*

[a]Department of Chemistry, Indian Institute of Technology Kanpur, Kanpur 208016, Uttar Pradesh, India
[b]Brahmananda Keshab chandra College, Kolkata-700108, West Bengal, India
[c]Centre for Laser and Photonics, Indian Institute of Technology Kanpur, Kanpur 208016, Uttar Pradesh, India

[¤]These authors contributed equally to this work
*dgoswami@iitk.ac.in

**Abstract:**

Microscopic thermodynamic studies can elucidate specific molecular interactions. In this work, we report the microscopic thermodynamics in binary liquid mixtures, which elucidate the role of molecular structure in nonlinear solvent response using femtosecond optical tweezers (FOT). We obtain the excess thermodynamics property of mixing in various Newtonian liquid mixtures by analyzing Microrheology data from FOT. Using our noninvasive 780 nm pulse laser we have trapped micron sized particles to show how excess viscosity and residual Gibbs free energy change due to mixing. Furthermore, we establish from this study that hydrocarbon chain length and branching can modulate microscopic thermodynamics through intermolecular interaction. This work sheds light on the relationship between thermodynamics properties and viscosity, which is of immense importance for predicting transport properties, mixing, and chemical reactions.

**Introduction**

The thermodynamic properties for a microscopic system, owing to their molecular interactions, can be different from the macroscopic ones. Ever growing interest in miniaturizing the existing technology calls for deeper understanding of the nature of thermodynamics at the microscopic level. There have been a good deal experimental and theoretical efforts devoted to understanding of nature of work in microscopic systems[1-4]. Excess viscosity and molar residual Gibbs energy of mixing for liquid mixtures are couple of thermodynamic properties associated with the microscopic architecture of the system, as they signify the deviation of the real system from the reference behavior [5]. In this paper, we have tried to understand how change in molecular structure is governed by the change in viscosity of the binary mixture as the relative concentration of the solvents is varied. Use of femtosecond optical tweezers (FOT) enable the measure of nonlinear solvent response that is critical to elucidate the role of molecular structure in viscosity variations that in turn enables us to analyze Microrheology. Water and alcohol binary mixtures are of immense importance in various fields of research, e.g., in studying the dynamics of protein folding[6], molecular segregation [7], enzyme reaction rates [8, 9] to name a few and have been extensively studied in the literature[6, 10-14]. The binary mixture of Newtonian fluids (e.g. alcohols) especially with water can create a model environment that supports most of the biochemical reactions to occur [15, 16]. As mentioned above, we use FOT to detect changes in rheology of fluids at microscopic level. Optical tweezers exploit the gradient force component of a strongly focused Gaussian beam to trap particles of micron to nm size [17] [18]. Trapped particles are suspended in a fluid and information of variation in the liquids viscosity in its vicinity gets encoded in particle's oscillation. This information encrypted in the particles motion is extracted with optical tweezers, which makes it an extremely sensitive technique for such studies. Over the period, optical tweezers have been extensively developed and have become an indispensable tool in numerous fields of research [19-26]. The non-invasiveness of selective trapping lasers has made it possible to study biological systems like cells, viruses, bacteria, etc. [27, 28] effortlessly. Probing micro or nano-rheology in complex media can easily be achieved by this technique [29-31]. We use power spectrum method [32] to calibrate our tweezers setup, it is based on the observation of the thermal motion of micron-sized spheres in low concentration (nanomolar), suspended in the viscoelastic system under investigation, assuming that the particles are non-interacting. For the present work, we have investigated mixtures of five different Newtonian fluids, with increasing carbon chain length at first and then with increase in branching. Variations in the micro-rheology of $H_2O$-MeOH, $H_2O$-EtOH, $H_2O$-nPrOH, $H_2O$-iPrOH, and $H_2O$-tBuOH at ten different volume proportions are experimentally determined and presented. Probing viscosity with nano size trapped bead of binary mixtures was carried out at a constant temperature and constant trapping power. The uniqueness of employing FOT technique to study micro-rheology of complex media as well as its excess properties is that it can sense the variations in fluid's rheology in a femtoliter volume.

**Experimental Methods and Materials**

In our femtosecond optical tweezers set-up, we have used a Ti:Sapphire (MIRA 900F) laser, pumped with Nd:YVO$_4$ (Verdi V-5), which produces femtosecond laser pulses centered at 780 nm wavelength with a pulse-width of 241.59 fs generated at a repetition rate of 76 MHz. Figure 1 is a simple schematic of our experimental setup; we control the laser power reaching the sample with a half-wave plate and polarizing beam-splitter placed in tandem. The beam then passes through a beam expander so that it has enough diameter to overfill the back aperture of the objective, in order to get the maximum gradient force possible. We used a commercial oil immersion objective (UPlanSApo, 100X, 1.4 NA, OLYMPUS Inc. Japan) to achieve tight focusing; simultaneously the forward scattered light was collected with another oil immersion objective (60x, PlanAapo N, 1.42 NA, OLYMPUS Inc. Japan) and focused onto a quadrant photodiode (QPD) (2901, Newport Co. USA) that has a risetime of 4μs and an active area of 3X3 mm. Using a torch, the sample can be viewed under white light illumination as well. Output from QPD is connected to a digital oscilloscope (Waverunner LT354M, LeCroy USA) interfaced with a personal computer through a GPIB card (National Instruments, USA).

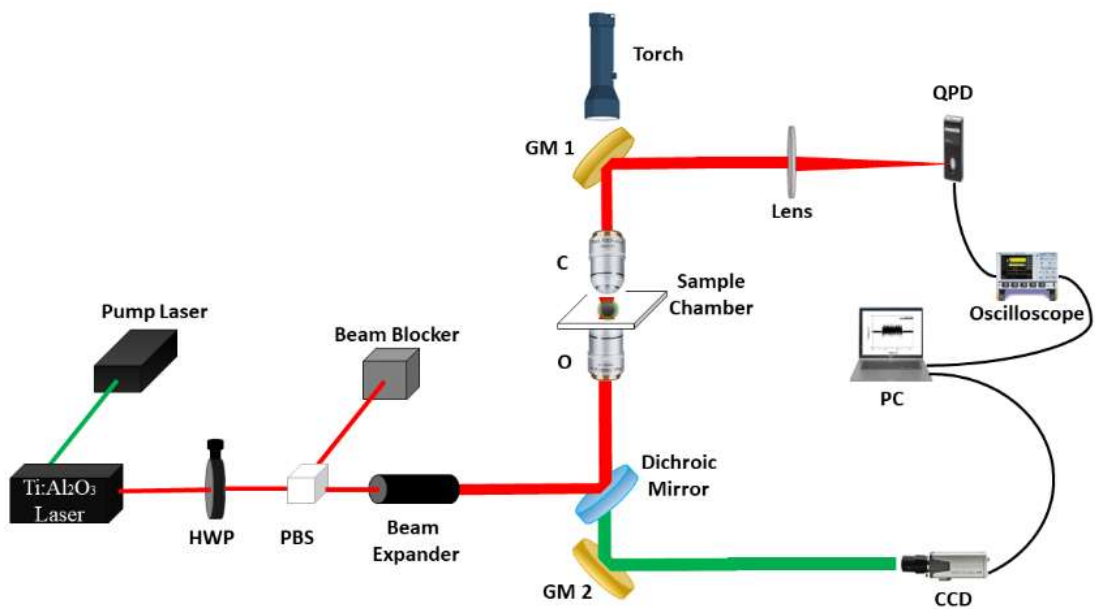

Figure 1 Optical tweezers setup with labels. HWP: Half waveplate; PBS: Polarizing beam-splitter; O: Objective lens; C: Condenser lens; GM: Gold mirrors; QPD: Quadrant photodiode; PC: Personal computer; CCD: Camera (charged coupled device).

Data was acquired at 100 kHz sampling rate using LabVIEW program. Video of the trapping events were monitored using CCD camera (350 K pixel, e-Marks Inc. USA). The trapping laser power was measured with a power meter (FieldMate, Coherent USA). Spectroscopic grade MeOH, EtOH, nPrOH, *i*PrOH, and tBuOH were purchased from Merck, India and were used without any further purification. We have used 24×50 mm No. 0 cover glass with an assembled by placing a coverslip 22×22 mm No. 0 separated by spacers of double-sided sticky tape. We used 500 nm mean radius (T8883, Life technology, USA) fluorophore coated polystyrene sphere and suspended them in $H_2O$-MeOH, $H_2O$-EtOH, $H_2O$-iPrOH and H2O-nPrOH mixture. The commercially available polystyrene nanosphere solution with concentration $3.6\times10^{10}$ particles/ml was diluted and well sonicated before performing trapping experiments. Figure 2 represents the QPD signal as seen on the oscilloscope, the voltage fluctuation signifies the Brownian motion of the trapped particle and the sharp spike at beginning is due to particle entering the trap.

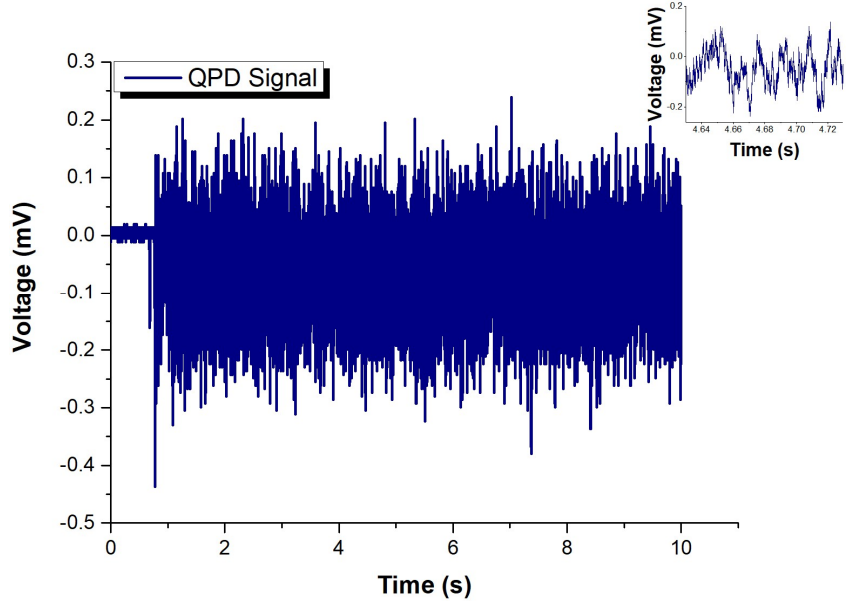

Figure 2 Typical QPD output showing voltage fluctuations when the particle gets trapped (inset shows zoomed in signal over a shorter time period).

## Theoretical Background

We have trapped 500 nm mean radius polystyrene beads with 780 nm femtosecond high repetition rate laser using our femtosecond optical tweezers set up fig1. The z direction stage is used to keep the focal plane of the trapping laser fixed from the surface of the cover glass for each of the solvents used that also verified with back scattered signal from our CCD camera. When nanometer size particle is in a viscous Newtonian fluid undergoing Brownian motion under oscillating harmonic potential well then equation of motion of that particle can be expressed as following Langevin equation under the approximation that the motion of nanometer-sized particles takes place at a small Reynolds number where viscous drag dominates inertial forces

$$\gamma \dot{x}(t) + \kappa x(t) = \zeta_{therm}(t) \tag{1}$$

Where x(t) is time dependent position, $\gamma$ is the viscous drag coefficient as per Stokes' Law, $\kappa$ is the spring constant and $\zeta_{therm} = \sqrt{(2k_BT/\gamma)}F(t) = \sqrt{(2D)}F(t)$ is the time dependent random thermal force. The diffusion coefficient can be expressed by Einstein equation $D = k_BT/\gamma$. By solving the above equation, we can fit our experimental power spectrum into theoretical power spectrum with the following Lorentzian [32]

$$p_x(f) = \frac{D}{2\pi^2(f_c^2 + f^2)} \tag{2}$$

However, in the power spectrum, power is measured in the interval f and f+df which does not distinguish between +f and –f. In such cases it is possible to define one sided power spectral density (PSD) as [33]

$$P_x(f) \equiv \frac{1}{T_{msr}}(|\tilde{x}(f)|^2 + |\tilde{x}(-f)|^2) \quad 0 \leq f < \frac{T_{msr}}{2}$$
$$= \frac{2}{T_{msr}}|\tilde{x}(f)|^2$$
$$= \frac{D}{\pi^2(f_c^2 + f^2)}$$
$$\tilde{x}_f(f) = \int_{-T_{msr}/2}^{T_{msr}/2} dt\, e^{i2\pi f_k} x(t), \quad f_k \equiv k/T_{msr}\}$$

We then analyze the data from QPD (fig 2(b)) with our MATLAB program and plot the corresponding one-sided power spectrum. The plotted power spectrum is then fitted with the Lorentzian curve and values of corner frequency ($f_c$) and diffusion coefficient (D) are obtained. We have used oil immersion objective with NA= $n_m \sin(\alpha)$ =1.4 with refractive index of the oil used is 1.518. The objective lens has a transmission of 65% at 780 nm wavelength and a working distance of 100 μm. The cover glass used has 80 μm width so in our set up particle will be trapped 20 μm above (h) from the cover glass surface. We have analyzed the first 5 seconds trapping data collected at 100 kHz sampling rate. The processed data was then fitted with the following Lorentzian function:

$$p_x^{(exp)}(f) = \frac{A}{(f_c^2 + f^2)} = \frac{D}{\pi^2(f_c^2 + f^2)} \tag{3}$$

The fitting parameter A is used to evaluate the diffusion coefficient D, introducing a new conversion factor α as the following way:

$$A(V^2/s) \times \alpha(nm^2 \cdot V^{-2}) = \frac{D}{\pi^2}(nm^2/s)$$

$$\pi^2 \times A \times \alpha = \frac{k_B T}{6\pi \eta r} \tag{4}$$

$$\eta = \frac{k_B T}{6\pi^3 \alpha r A} \tag{5}$$

Here α introduced is a conversion factor from voltage to position. At room temperature we can easily evaluate the value α without performing any additional voltage calibration experiment. As we have used the same trapping power and the refractive indices vary with 5% error bar, we can use the same calibration factor for each measurement for the same set of experiment.

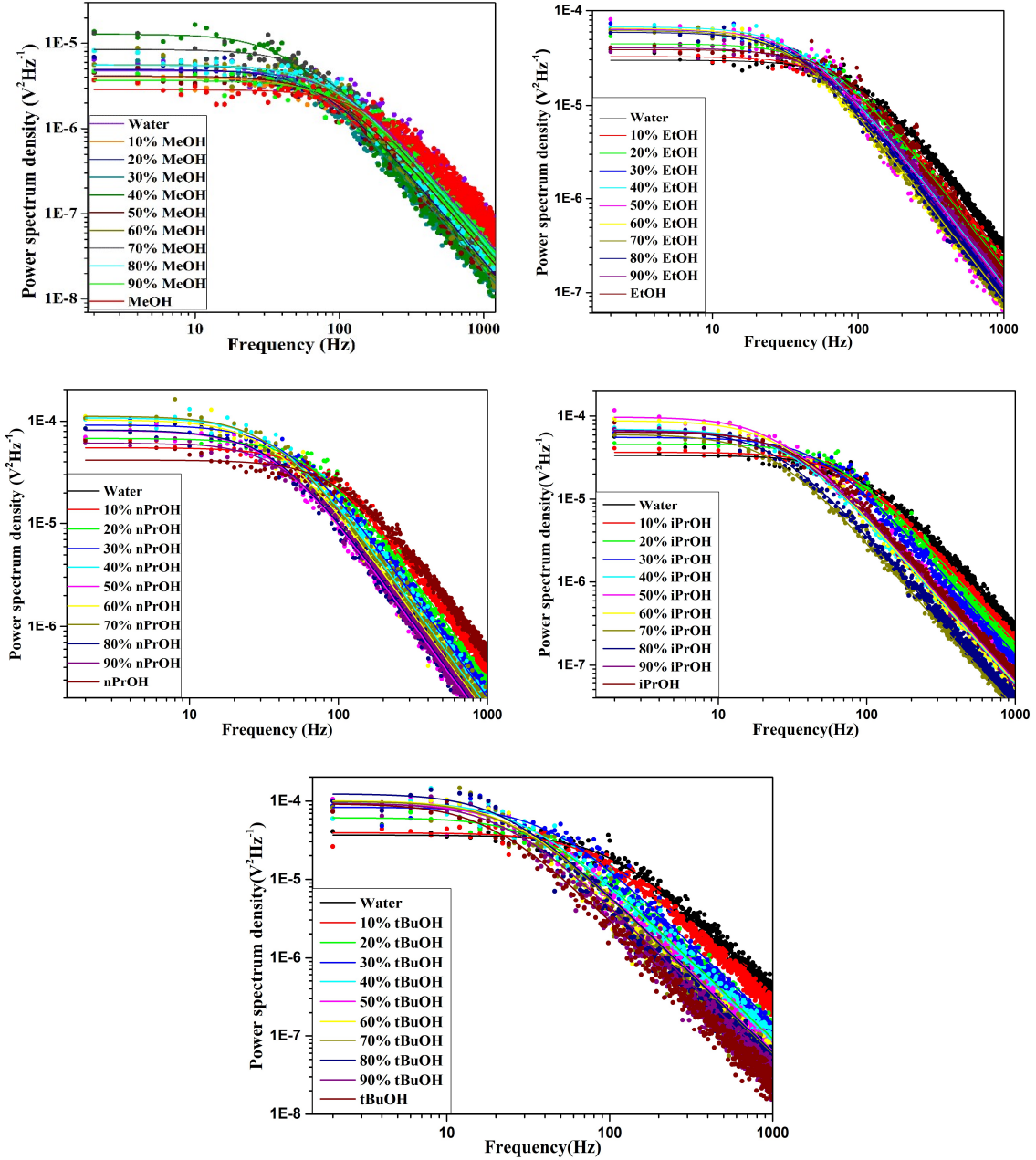

Figure 3 Power spectrum for water-alcohol binary mixtures with their corresponding Lorentzian fit.

The Viscosity of non-ideal binary mixtures is considered as by the following equation [5, 34]

$$\left.\begin{array}{l} ln\eta^{ex} = ln\,\eta_\varphi^{id} + \Delta\,ln\,\eta_\varphi^{ex} \\ = (\varphi_1\,ln\,\eta_1 + \varphi_2\,ln\,\eta_2) + \Delta\,ln\,\eta_\varphi^{ex} \end{array}\right\} \quad (6)$$

where φ and η are the volume fraction and viscosity of the pure substances respectively and indices 1 and 2 stands for component 1 and 2 of the binary mixtures. For our calculations we have reserved index 1 for water and index 2 for alcohols. Since the interaction between two individual solvent molecules in their homogeneous binary mixture can considerably alter the viscosity of the medium. The Δln(η$^{ex}$) term is called excess viscosity which is indicative of the deviations of real system from ideal one, it can be evaluated theoretically as well as experimentally. In this article we have experimentally measured the viscosity of $H_2O$-MeOH, $H_2O$-EtOH, $H_2O$-nPrOH, $H_2O$-$i$PrOH and $H_2O$-$t$BuOH. Wolf et. al suggested that theoretical excess viscosity can be calculated using the following equation:

$$\Delta \ln \eta^{theo} = \frac{\gamma \delta}{(1+\gamma \varphi_2)} \varphi_2 (1-\varphi_2) + 2g \frac{(1+\gamma)}{(1+\gamma \varphi_2)^2} \varphi_2 (1-\varphi_2) \quad (7)$$

The significant assumption of this model is that the frictional force works between two molecules mainly depends on their interfaces rather their volume so the difference between the surface fraction Ω (Ω=(1+γ)φ/(1+φγ)) and the volume fraction φ can be measured by the geometric factor $\gamma = \frac{F_2/V_2}{F_1/V_1} - 1$. Here F and V are represented as molecular surface and molar volume respectively. Commonly the first term in equation (7) is mostly dominating and the second term comes in play only for systems with significant interactions. The δ can be calculated using relation δ = lnη$_2$−lnη$_1$. The temperature dependent g, is the interaction parameter can be calculated from the following equation $g = \frac{\Delta G^{Rd}}{RTx_1\varphi_2}$ where $x_1$ is the mole fraction of component 1 and $\varphi_2$ is the volume fraction of component 2 of the binary mixtures and $\Delta G^R$ is molar residual Gibbs free energy of mixing at temperature T in K. Trap stiffness (κ = 2πf$_c$γ) is preferentially considered as a constant parameter due to continuous change in viscosity with heating laser source power within same solvent media for a constant sized trapped particle.

**Results and Discussions**

For the present work we have chosen methanol, ethanol, propanol, iso-propanol and *t*-butanol as they are the most important of the amphiphilic solvents which have both hydrophobic and hydrophilic groups which permits them to interact with large molecules and biopolymers in many different ways, they can also form aggregates in water while remaining soluble, which make small pockets of heterogenous environment in the solution[35]. There are two effects which are simultaneously at play, there is formation of hydrogen bonds between the hydrophilic part and water molecules and the hydrophobic hydration effect which may induce cooperative ordering in the system due to the presence of hydrophobic part. The combined outcome of these effects governs the extent of hydrogen bonding network of water in the binary mixtures is apparent in the non-ideal behavior of many physical properties such as viscosity, density, excess mixing volume, etc. Previous viscosity studies for water-MeOH, water-EtOH and water-*n*PrOH [36] using rheometer shows a monotonous variation in the viscosity of binary mixtures and nowhere over the entire composition range do the curves for individual binary mixtures cross each other. With our FOT setup we have looked at the non-ideal variation in the excess viscosity of these binary mixtures, first based on the increase in carbon chain length and then by increase in the branching network of the alcohol molecules. We extract geometric factor γ that accounts for the effects that arise due to differences in the size of molecules of identical shape and differences in the molecular structure, and the thermodynamic interaction parameter g, which contains information related to the intermolecular interaction between the molecules of two solvents. For binary fluids with g greater than 1 is indicative of the significant interaction between the molecules present in the fluid and thus makes the second term in equation 7 consequential. To determine microscopic thermodynamic parameters i.e., excess viscosity and Gibbs energy of mixing associated with the binary mixture, we first looked at the variation in the diffusion coefficient that we get from fitting the experimental power spectrum. Diffusion coefficient associated with the trapped bead is our first insight into the changing behavior of the binary mixture due to change in relative concentration of the individual solvents. Variation in the diffusion coefficient for the binary mixtures, based on increasing carbon chain length and branching is presented in figure 4 (a) and figure 4 (b) respectively (refer to the supplementary information for specific values). Both the graphs have the same origin point as it represents the diffusion coefficient for water and as the relative volume of alcohol is increased in the binary mixture, the diffusion coefficient for water-MeOH binary mixture decreases initially up to 0.3 volume fraction and then goes through a sudden jump from 0.4 to 0.5 volume fraction, which is indicative of the drastic change in the micro-environment between these concentrations, and beyond these the trend is linear towards pure MeOH character. In water-EtOH binary system there is sharp change in the diffusion coefficient from 0.2 to 0.3 volume fraction of EtOH, and in water-*n*PrOH binary system the anomaly occurs at 0.5 volume fraction of *n*PrOH.

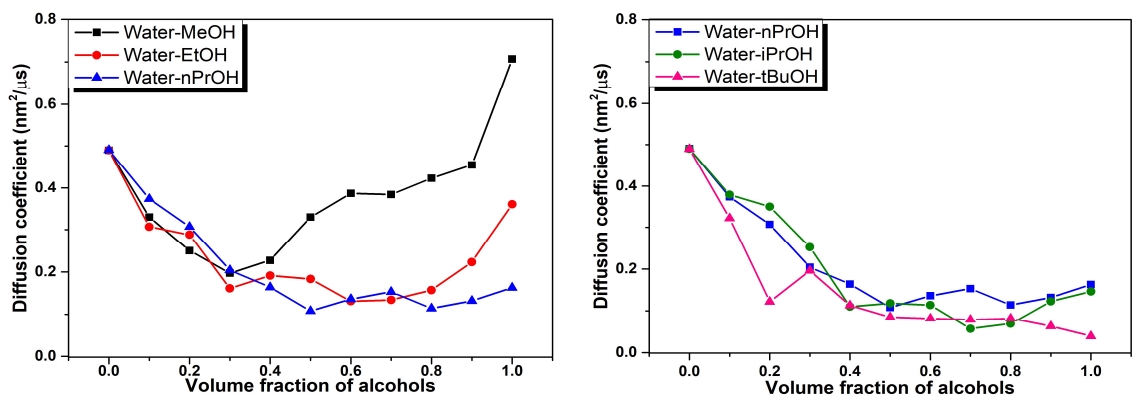

Figure 4 Variation in the diffusion coefficient(a) represents change due to increase in carbon chain length (b) changes due to increase in branching in alcohols.

Whereas, in case of branched systems, for water-*i*PrOH the anomaly in the diffusion coefficient trend seems to occur at 0.4 and 0.7 volume fraction. And finally in water-*t*BuOH system 0.2 volume fraction of *t*BuOH appears to be the region where significant change in the micro-environment takes place. With the knowledge of variation in the diffusion coefficient, we can then use equation 5 and calculate the variation in the viscosity for each of the binary system. Figure 5 (a) and (b) correspond to variation in the experimentally observed viscosity and ideal viscosity of water-MeOH, water-EtOH and water-*n*PrOH respectively (refer to the supplementary information for specific values). Where we have calculated the ideal viscosities from equation 6. The viscosity variation of water-MeOH binary system shown in figure 5 (a) shows a peak at 0.3 and 0.4 volume fraction and thereafter a smooth decay towards pure MeOH viscosity. The difference in viscosities at 0.3 and 0.4 volume fraction from the rest of the binary mixtures in the series could be attributed to the difference in the hydrogen bonding network at these concentrations. For water-EtOH, there are sharp peaks in the viscosity curve at 0.3 and 0.6 volume fraction suggesting different hydrogen bonding network in the binary mixture. Whereas, in water-*n*PrOH viscosity curve the peak suggesting anomalies occur at 0.5 and 0.8 volume fractions. On the left figure 5 (b) shows how the viscosities in an ideal binary system for the would have behaved if not for the intermolecular interaction.

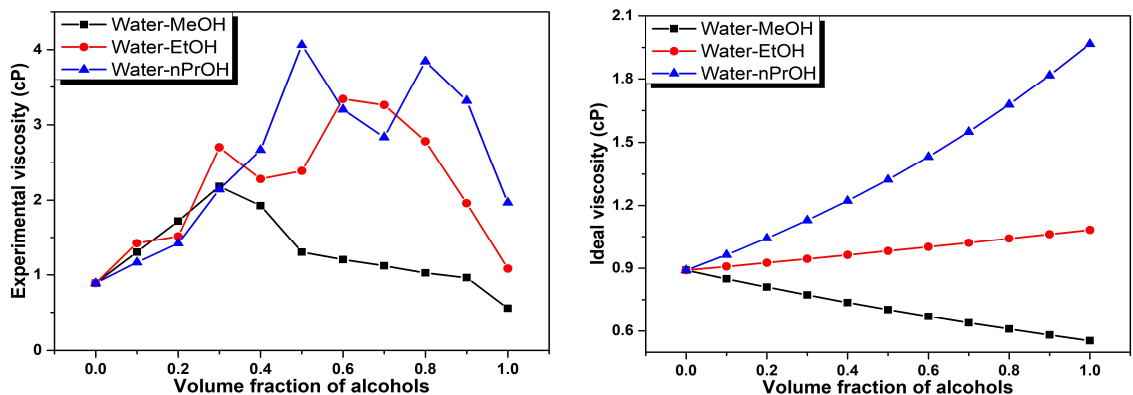

Figure 5 (a) and (b) represent the variation in the experimental and ideal viscosity for water-MeOH, water-EtOH and water-*n*PrOH binary systems.

Figure 6 (a) and (b) represent the change in the experimental and ideal viscosity of $n$PrOH, $i$PrOH and $t$BuOH respectively (refer to the supplementary information for specific values) as the branching in the alcohol is increased with respect to $n$PrOH. On examining water-$i$PrOH viscosity curve, we see two significant peaks at 0.4 and 0.7 volume fractions and the peak at 0.7 volume fraction is significantly higher than the rest of the binary mixtures in the water-$i$PrOH series, which means the interaction between water and $i$PrOH molecules is quite different at this particular concentration. And in water-$t$BuOH series the peaks appear at 0.2 and 0.8 volume fraction. The viscosity curves plotted in figure 6 (b) uses equation 6 to predict the ideal variation in the viscosities of water-$i$PrOH and water-$t$BuOH binary mixtures.

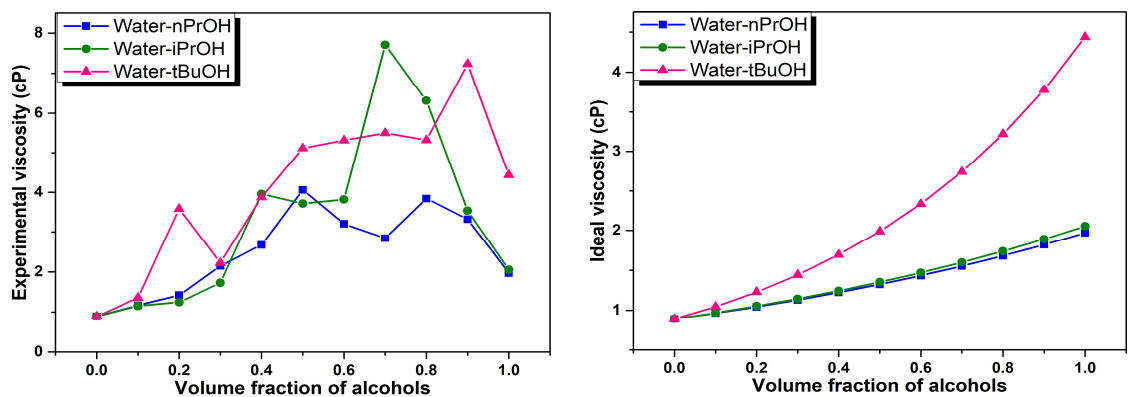

Figure 6 (a) and (b) represent the variation in the experimental and ideal viscosities of the binary mixtures respectively.

It is evident on comparison of figures 5 and 6, that variation in the experimental viscosity is very different from the ideality. And it is this deviation that we quantify by evaluating excess viscosity. Therefore, of the two microscopic thermodynamic parameters that we set out to measure; we first determine the excess viscosity. The proportion of the individual molecules in a Newtonian fluid dictates the viscosity of the media and by determining the excess viscosities we have commented on the molecular architecture of the respective fluids. We fitted the plot of variation of excess viscosity with change in volume fraction for individual binary mixture with equation 7 and evaluated the geometric factor $\gamma$ and thermodynamic interaction parameter g (specific figures with their corresponding fitting curves can be found in the supplementary information). Of the linear chain systems, water-EtOH had the maximum g value of 2.40, in comparison to 1.71 for water-MeOH and 1.89 for water-$n$PrOH. And of the branched systems water-$i$PrOH had the higher g value of 2.68 than 1.80 for the water-$t$BuOH case. For all the cases, g values are suggestive of significant interaction among the solvent molecules, and it is probably the main reason why the variation in experimental viscosity is nonlinear with increasing alcohol concentration. It is also evident from figure 7 (a) that at 0.3 volume fraction for water-MeOH system, at 0.6 volume fraction for water-EtOH and at 0.5 volume fraction for water-$n$PrOH, there is maximum deviation from ideality. Whereas from figure 7 (b), it is clear that at 0.7 volume fraction for water-$i$PrOH and at 0.2 volume fraction for water-$t$BuOH is where maximum deviation occurs. Since all the alcohols used in present study are amphiphilic in nature and it is known that although these amphiphilic cosolvents do form binary mixture that would remain soluble for the entire composition range, their chemical composition at the micron scale may very well be different from the bulk[35]. And the nonlinear variation in the experimental viscosity for the binary mixtures suggests that we can sense these micro-heterogeneous environments with femtosecond optical tweezers.

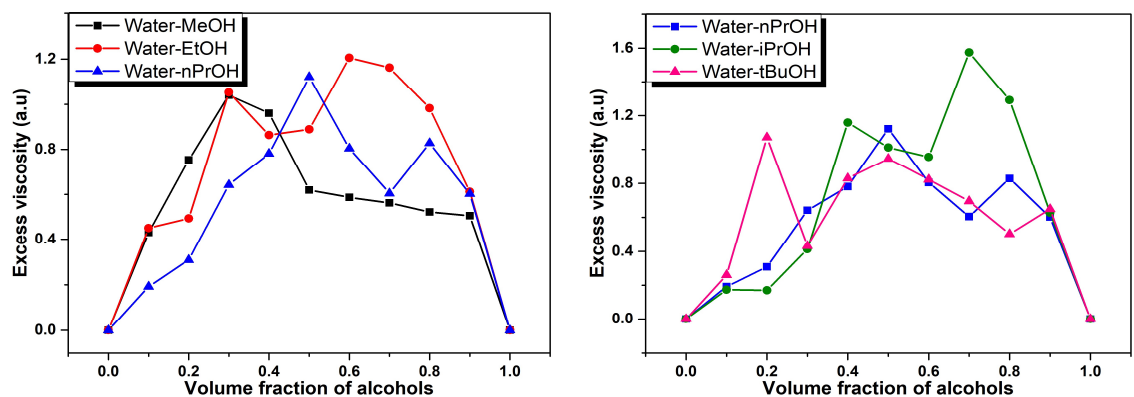

Figure 7 Variation in excess viscosity with (a) increasing carbon chain length and (b) increasing branching.

Now coming onto the second excess property, as residual properties represent the difference between a certain thermodynamic property in ideal state versus when it is in a real state, by deducing the variation in the molar residual Gibbs energy of mixing for individual system we can further comment on the interaction in the respective water-alcohol binary mixture. From the relationship between thermodynamic interaction parameter g and molar residual Gibbs energy of mixing, $\Delta G^R$, we calculated respective values of $\Delta G^R$ at all the concentrations (refer to the supplementary information for specific values) and represented them in figure 8 (a) and (b). The curves for $\Delta G^R$ exhibit smooth variation in the values because the volume fraction of component 1 (water) and mole fraction of component 2 (alcohol) is varied linearly. Among the straight chained alcohols, water-EtOH binary system had the maximum g value of 2.40 and therefore its corresponding curve in figure 8 (a) lies above both water-MeOH (g=1.71) and water-$n$ProH (g=1.89) systems which have relatively smaller g values. Similarly for binary system of water and branched alcohols, water-$i$PrOH binary system had the g value of 2.67 and therefore has higher $\Delta G^R$ values than the other two at all the concentrations. Whereas water-$n$PrOH and water-$t$BuOH had extremely close g values and thus their respective curves in figure 8 (b) overlap each other for almost all the concentrations.

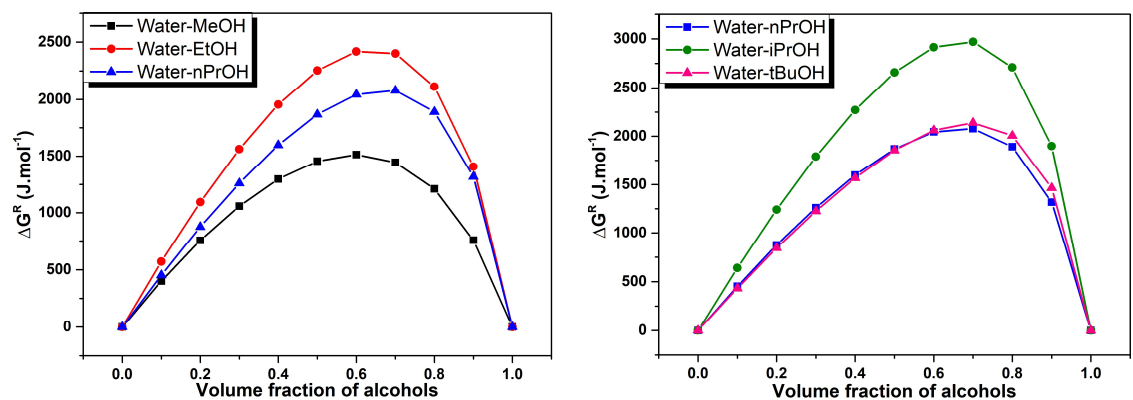

Figure 8 Variation in the residual Gibbs energy of mixing for water-alcohol binary mixtures

## Conclusion

We have investigated five water-alcohol binary mixture systems (Water-MeOH, Water-EtOH, Water-$n$PrOH, Water-$i$PrOH and Water-$t$BuOH) and the excess property of mixing is extensively studied by probing nano-size trapped particle under femtosecond laser. The dynamic viscosities within incompressible solvent mixtures are measured and characterized. The role of molecular structure in nonlinear solvent response was possible due to the use of the FOT.

The cross-over points in the curves are the regions of non-ideality. The measure of excess thermodynamic properties at microscale dimension leads us to comment on the intermolecular interaction between the amphiphilic alcohol and water molecules around the trapped bead. All five systems exhibit relatively higher values for the thermodynamic interaction parameter g which suggests that these binary mixtures have significant intermolecular interaction and the sudden change in viscosities at certain concentrations is indicative of the relative change in the molecular arrangement at those corresponding concentrations. Water-EtOH and water-*i*PrOH systems had the maximum value for the thermodynamic interaction parameter suggesting higher extent of interaction among the molecules of those binary mixtures.

# Supplementary Information

a) **Water-MeOH system:**

| % Of MeOH | Mole Fraction of MeOH | Volume fraction | Diffusion coefficient (nm$^2$/µs) | Experimental Viscosity (cP) | Ideal Viscosity (cP) | Excess Viscosity (a.u.) | $\Delta G^R$ (J.mol$^{-1}$) |
|---|---|---|---|---|---|---|---|
| Pure water | 0 | 0 | 0.490 | 0.8900 | 0.8900 | 0 | 0 |
| 10% MeOH | 0.0472 | 0.1 | 0.330 | 1.3205 | 0.8487 | 0.4420 | 404.48 |
| 20% MeOH | 0.1004 | 0.2 | 0.250 | 1.7450 | 0.8094 | 0.7682 | 764.13 |
| 30% MeOH | 0.1606 | 0.3 | 0.197 | 2.2109 | 0.7720 | **1.0521** | 1069.95 |
| 40% MeOH | 0.2294 | 0.4 | 0.227 | 1.9236 | 0.7362 | 0.9604 | 1310.38 |
| 50% MeOH | 0.3087 | 0.5 | 0.330 | 1.3241 | 0.7021 | 0.6344 | 1470.28 |
| 60% MeOH | 0.4011 | 0.6 | 0.387 | 1.1284 | 0.6696 | 0.5218 | 1529.46 |
| 70% MeOH | 0.5103 | 0.7 | 0.384 | 1.1363 | 0.6386 | 0.5762 | 1460.36 |
| 80% MeOH | 0.6411 | 0.8 | 0.423 | 1.0308 | 0.6090 | 0.5262 | 1224.34 |
| 90% MeOH | 0.8007 | 0.9 | 0.455 | 0.9599 | 0.5808 | 0.5024 | 765.53 |
| Pure MeOH | 1 | 1 | 0.707 | 0.5540 | 0.5540 | 0 | 0 |

*Table 1 Experimental parameters for Water-MeOH binary mixtures.*

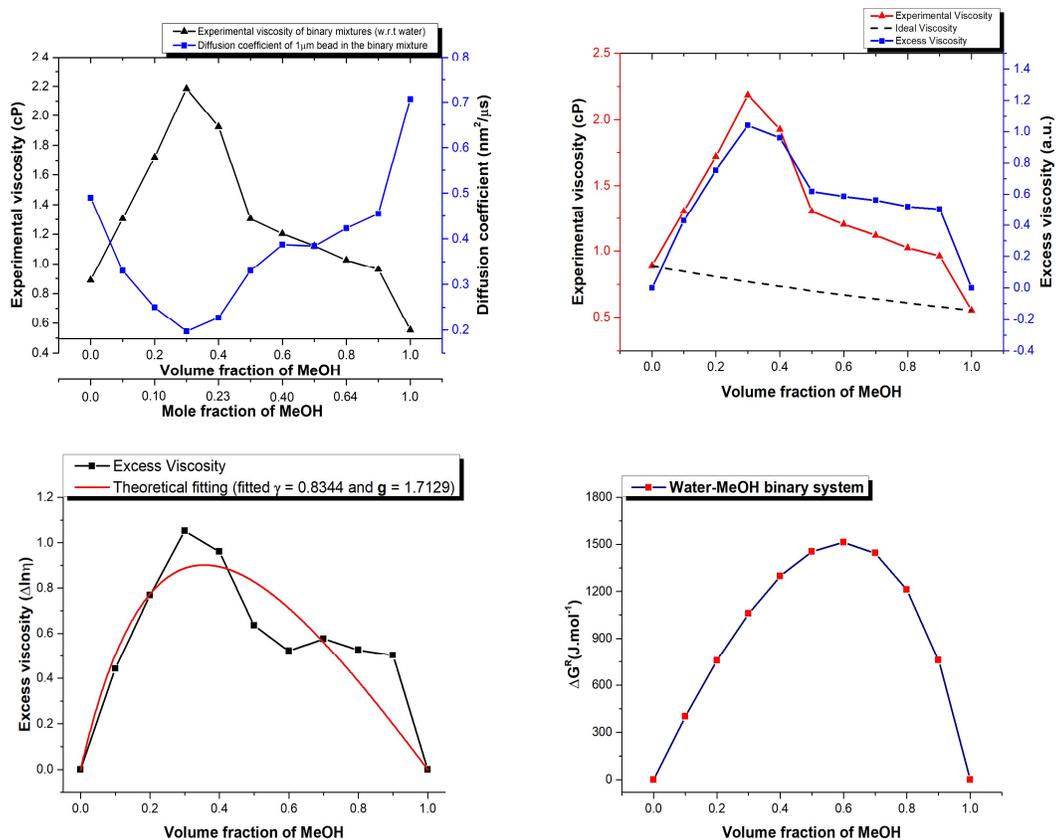

*Figure 1 (a) Plot of experimental viscosity and diffusion coefficient vs volume fraction of MeOH. (b) Plot of experimental viscosity and excess viscosity vs volume fraction of MeOH. (c) Excess viscosity variation with volume fraction of MeOH (with fitted values of γ and g in inset) (d) Variation of ΔG$^R$ with MeOH volume fraction*

b) **Water-EtOH system:**

| % Of EtOH | Mole fraction of EtOH | Volume fraction | Diffusion coefficient (nm$^2$/μs) | Experimental Viscosity (cP) | Ideal Viscosity (cP) | Excess Viscosity (a.u.) | ΔG$^R$ (J.mol$^{-1}$) |
|---|---|---|---|---|---|---|---|
| Pure Water | 0 | 0 | 0.490 | 0.8900 | 0.8900 | 0 | 0 |
| 10% EtOH | 0.0332 | 0.1 | 0.307 | 1.4233 | 0.9076 | 0.4499 | 575.57 |
| 20% EtOH | 0.0717 | 0.2 | 0.288 | 1.5147 | 0.9256 | 0.4925 | 1105.50 |
| 30% EtOH | 0.1170 | 0.3 | 0.161 | 2.7055 | 0.9439 | 1.0530 | 1577.80 |
| 40% EtOH | 0.1709 | 0.4 | 0.191 | 2.2838 | 0.9626 | 0.8639 | 1975.91 |
| 50% EtOH | 0.2362 | 0.5 | 0.183 | 2.3897 | 0.9817 | 0.8896 | 2276.26 |
| 60% EtOH | 0.3169 | 0.6 | 0.131 | 3.3421 | 1.0012 | **1.2054** | 2444.11 |
| 70% EtOH | 0.4192 | 0.7 | 0.134 | 3.2638 | 1.0210 | 1.1621 | 2426.03 |
| 80% EtOH | 0.5530 | 0.8 | 0.157 | 2.7840 | 1.0413 | 0.9834 | 2135.40 |
| 90% EtOH | 0.7357 | 0.9 | 0.223 | 1.9578 | 1.0619 | 0.6117 | 1421.95 |
| Pure EtOH | 1 | 1 | 0.361 | 1.0830 | 1.0830 | 0 | 0 |

*Table 2 Experimental parameters for Water-EtOH binary mixtures.*

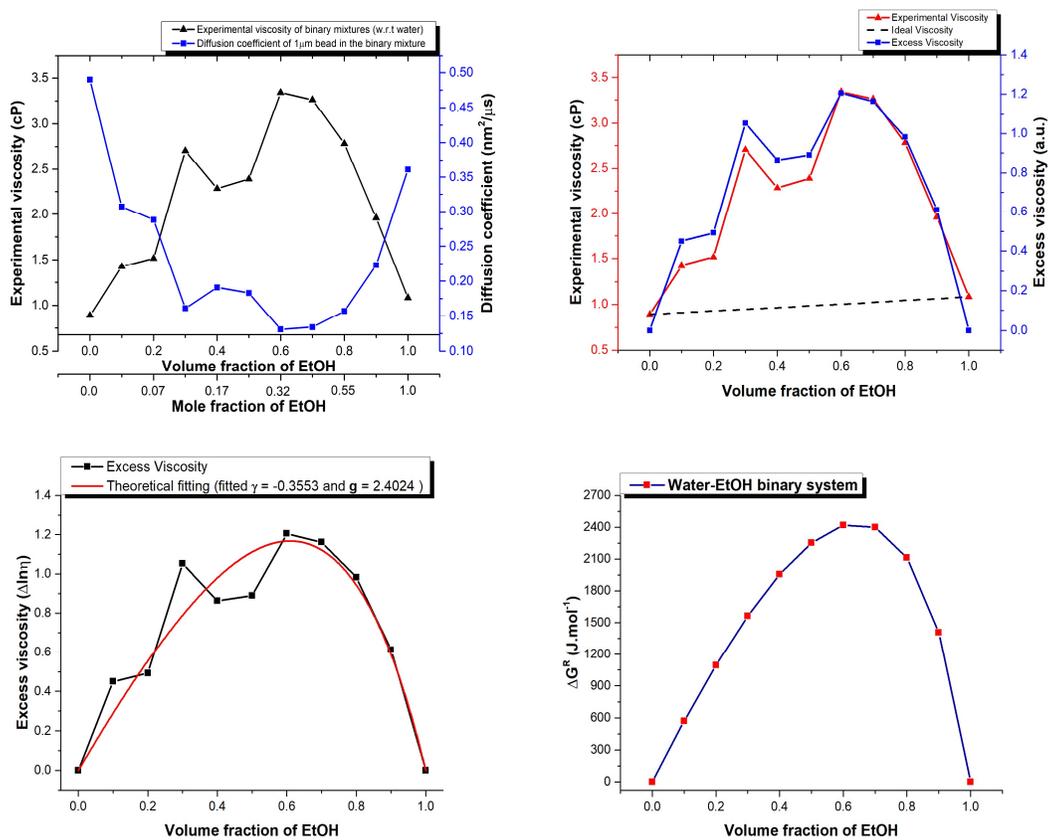

*Figure 2 (a) Plot of experimental viscosity and diffusion coefficient vs volume fraction of EtOH. (b) Plot of experimental viscosity and excess viscosity vs volume fraction of EtOH. (c) Excess viscosity variation with volume fraction of EtOH (with fitted values of γ and g in the inset)*

### c) Water-nPrOH system:

| % Of nPrOH | Mole fraction of nPrOH | Volume fraction | Diffusion coefficient (nm$^2$/μs) | Experimental Viscosity (cP) | Ideal Viscosity (cP) | Excess Viscosity (a.u.) | $\Delta G^R$ (J.mol$^{-1}$) |
|---|---|---|---|---|---|---|---|
| Pure Water | 0 | 0 | 0.491 | 0.8900 | 0.8900 | 0 | 0 |
| 10% nPrOH | 0.0261 | 0.1 | 0.374 | 1.1673 | 0.9634 | 0.1919 | 456.07 |
| 20% nPrOH | 0.0569 | 0.2 | 0.307 | 1.4210 | 1.0429 | 0.3093 | 883.41 |
| 30% nPrOH | 0.0937 | 0.3 | 0.204 | 2.1457 | 1.1290 | 0.6421 | 1273.54 |
| 40% nPrOH | 0.1386 | 0.4 | 0.164 | 2.6705 | 1.2222 | 0.7816 | 1614.28 |
| 50% nPrOH | 0.1944 | 0.5 | 0.108 | 4.0621 | 1.3231 | **1.1217** | 1887.47 |

| 60% nPrOH | 0.2658 | 0.6 | 0.136 | 3.2035 | 1.4323 | 0.8049 | 2064.84 |
| 70% nPrOH | 0.3603 | 0.7 | 0.153 | 2.8384 | 1.5505 | 0.6046 | 2099.77 |
| 80% nPrOH | 0.4912 | 0.8 | 0.114 | 3.8425 | 1.6785 | 0.8282 | 1909.53 |
| 90% nPrOH | 0.6848 | 0.9 | 0.132 | 3.3188 | 1.8170 | 0.6024 | 1331.96 |
| Pure nPrOH | 1 | 1 | 0.163 | 1.9670 | 1.9670 | 0 | 0 |

*Table 3 Experimental parameters for Water-nPrOH binary mixtures.*

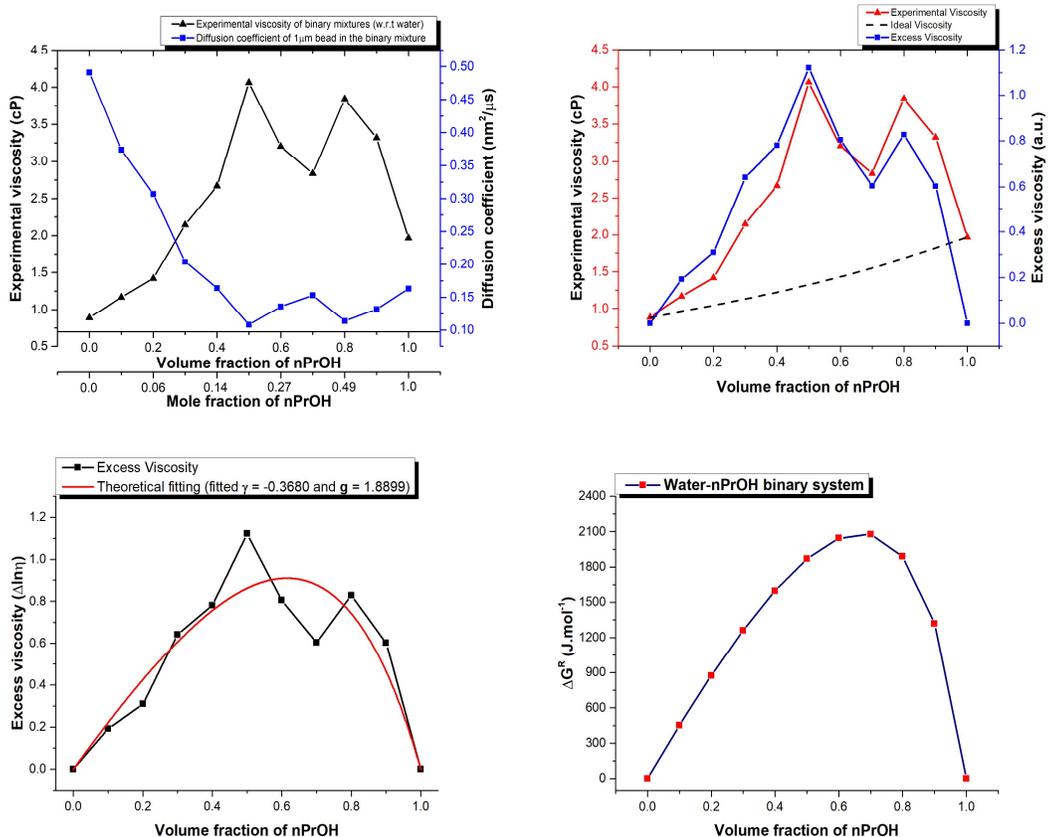

*Figure 3 (a) Plot of experimental viscosity and diffusion coefficient vs volume fraction of nPrOH. (b) Plot of Experimental viscosity and excess viscosity vs volume fraction of nPrOH. (c) Variation of Excess viscosity with volume fraction of nPrOH (with fitted values of γ and g in the inset). (d) Variation in residual Gibbs energy of mixing with volume fraction of nPrOH.*

d) **Water-*i*PrOH system:**

| % Of iPrOH | Mole fraction of iPrOH | Volume fraction | Diffusion coefficient (nm²/μs) | Experimental Viscosity (cP) | Ideal Viscosity (cP) | Excess Viscosity (a.u.) | $\Delta G^R$ (J.mol$^{-1}$) |
|---|---|---|---|---|---|---|---|
| Pure Water | 0 | 0 | 0.490 | 0.8900 | 0.8900 | 0 | 0 |
| 10% iPrOH | 0.0255 | 0.1 | 0.379 | 1.1513 | 0.9675 | 0.1739 | 646.62 |
| 20% iPrOH | 0.0557 | 0.2 | 0.350 | 1.2468 | 1.0518 | 0.1700 | 1253.40 |
| 30% iPrOH | 0.0919 | 0.3 | 0.252 | 1.7299 | 1.1434 | 0.4140 | 1808.44 |
| 40% iPrOH | 0.1360 | 0.4 | 0.110 | 3.9596 | 1.2430 | 1.1586 | 2294.65 |
| 50% iPrOH | 0.1911 | 0.5 | 0.118 | 3.7149 | 1.3513 | 1.0112 | 2686.44 |
| 60% iPrOH | 0.2616 | 0.6 | 0.114 | 3.8178 | 1.4691 | 0.9550 | 2943.74 |
| 70% iPrOH | 0.3553 | 0.7 | 0.057 | 7.7109 | 1.5971 | **1.5744** | 3000.09 |
| 80% iPrOH | 0.4858 | 0.8 | 0.069 | 6.3184 | 1.7362 | 1.2917 | 2736.59 |
| 90% iPrOH | 0.6801 | 0.9 | 0.123 | 3.5356 | 1.8875 | 0.6276 | 1917.51 |
| Pure iPrOH | 1 | 1 | 0.146 | 2.0520 | 2.0520 | 0 | 0 |

*Table 4 Experimental parameters for water-iPrOH binary mixtures.*

*Figure 4 (a) Plot of experimental viscosity and diffusion coefficient vs volume fraction of iPrOH. (b) Plot of Experimental viscosity and excess viscosity vs volume fraction of iPrOH. (c) Variation of Excess viscosity with volume fraction of iPrOH (with values of $\gamma$ and g in the inset). (d) Variation in the $\Delta G^R$ with volume fraction of iPrOH.*

e) **Water-*t*BuOH system**

| % Of tBuOH | Mole fraction of tBuOH | Volume fraction | Diffusion coefficient (nm²/μs) | Experimental Viscosity (cP) | Ideal Viscosity (cP) | Excess Viscosity (a.u.) | $\Delta G^R$ (J.mol$^{-1}$) |
|---|---|---|---|---|---|---|---|
| | | | | | | | |
| Pure Water | 0 | 0 | 0.490 | 0.8900 | 0.8900 | 0 | 0 |
| 10% tBuOH | 0.0207 | 0.1 | 0.322 | 1.3566 | 1.0451 | 0.2608 | 436.79 |
| 20% tBuOH | 0.0454 | 0.2 | 0.122 | 3.5822 | 1.2273 | **1.0711** | 851.53 |
| 30% tBuOH | 0.0754 | 0.3 | 0.196 | 2.2232 | 1.4413 | 0.4334 | 1237.13 |
| 40% tBuOH | 0.1126 | 0.4 | 0.113 | 3.8769 | 1.6925 | 0.8288 | 1583.14 |
| 50% tBuOH | 0.1599 | 0.5 | 0.085 | 5.1151 | 1.9876 | 0.9452 | 1873.40 |
| 60% tBuOH | 0.2221 | 0.6 | 0.082 | 5.3128 | 2.3341 | 0.8224 | 2081.57 |
| 70% tBuOH | 0.3076 | 0.7 | 0.079 | 5.4988 | 2.7410 | 0.6962 | 2161.66 |
| 80% tBuOH | 0.4323 | 0.8 | 0.082 | 5.3214 | 3.2188 | 0.5027 | 2025.38 |
| 90% tBuOH | 0.6314 | 0.9 | 0.063 | 7.2352 | 3.7800 | 0.6492 | 1479.10 |
| Pure tBuOH | 1 | 1 | 0.039 | 4.4390 | 4.4390 | 0 | 0 |

*Table 5 Experimental parameters for Water-tBuOH binary mixtures.*

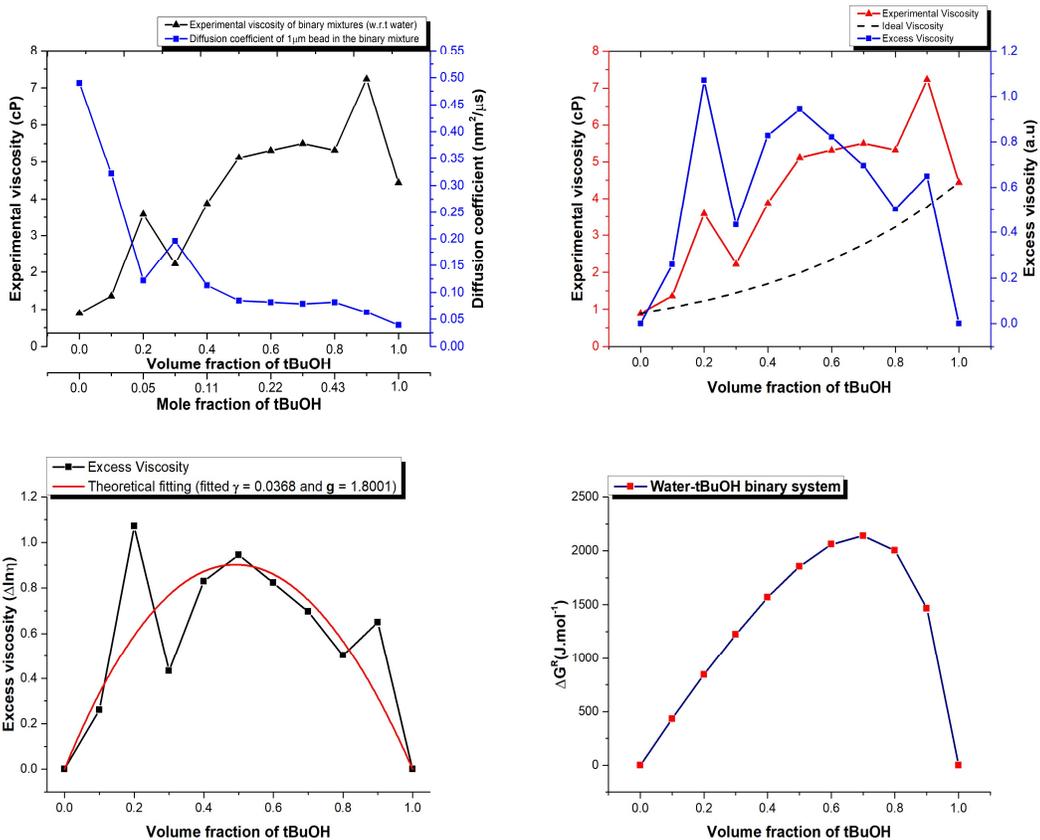

*Figure 5 (a) Plot of experimental viscosity and diffusion coefficient vs volume fraction of tBuOH. (b) Plot of Experimental viscosity and excess viscosity vs volume fraction of tBuOH, (c) Variation of ΔG$^R$ with volume fraction of alcohols. (d) Variation in the ΔG$^R$ with volume fraction of tBuOH.*